\documentclass[aps,pre,twocolumn,superscriptaddress]{revtex4}
\usepackage{graphicx}
\usepackage{dcolumn}
\usepackage{color}
\usepackage{amsmath}
\usepackage{amssymb}
\usepackage{appendix}

\begin{document}

\title{High-order harmonic generation in an electron-positron-ion plasma}

\author{W. L. Zhang}
\email[Corresponding author, ]{wenlong.zhang@tecnico.ulisboa.pt}
\affiliation{GoLP/Instituto de Plasmas e Fusão Nuclear, Instituto Superior Técnico, Universidade de Lisboa, Lisboa, Portugal}

\author{T. Grismayer}
\affiliation{GoLP/Instituto de Plasmas e Fusão Nuclear, Instituto Superior Técnico, Universidade de Lisboa, Lisboa, Portugal}

\author{K. M. Schoeffler}
\affiliation{GoLP/Instituto de Plasmas e Fusão Nuclear, Instituto Superior Técnico, Universidade de Lisboa, Lisboa, Portugal}

\author{R. A. Fonseca}
\affiliation{GoLP/Instituto de Plasmas e Fusão Nuclear, Instituto Superior Técnico, Universidade de Lisboa, Lisboa, Portugal}
\affiliation{DCTI/ISCTE—Instituto Universitário de Lisboa, 1649-026 Lisboa, Portugal}

\author{L. O. Silva}
\email[Corresponding author, ]{luis.silva@tecnico.ulisboa.pt}
\affiliation{GoLP/Instituto de Plasmas e Fusão Nuclear, Instituto Superior Técnico, Universidade de Lisboa, Lisboa, Portugal}

\date{\today}

\begin{abstract}
The laser interaction with an electron-positron-ion mixed plasma is studied, from the perspective of the associated high-order harmonic generation. For an idealized mixed plasma which is assumed with a sharp plasma-vacuum interface and uniform density distribution, when it is irradiated by a weakly relativistic laser pulse, well-defined signals at harmonics of the plasma frequency in the harmonic spectrum are observed. These characteristic signals are attributed to the inverse two-plasmon decay of the counterpropagating monochromatic plasma waves which are excited by the energetic electrons and the positron beam accelerated by the laser. Particle-in-cell simulations show the signal at twice the plasma frequency can be observed for a pair density as low as $\sim 10^{-5}$ of the plasma density. In the self-consistent scenario of pair production by an ultraintense laser striking a solid target, particle-in-cell simulations, which account for quantum electrodynamic effects (photon emission and pair production), show that dense (greater than the relativistically-corrected critical density) and hot pair plasmas can be created. The harmonic spectrum shows weak low order harmonics, indicating a high laser absorption due to quantum electrodynamic effects. The characteristic signals at harmonics of the plasma frequency are absent, because broadband plasma waves are excited due to the high plasma inhomogeneity introduced by the interaction. However, the high frequency harmonics are enhanced due to the high-frequency modulations from the direct laser coupling with created pair plasmas.
\end{abstract}


\maketitle
\section{Introduction}
\label{Sec: introduction}
Electron-positron pair plasmas play a key role in the exotic physics associated with astrophysical objects in extreme (ultra-massive, ultra-relativistic, or extreme magnetic fields, etc.) conditions \cite{Ruffini2010,Uzdensky2011}, such as active galactic nuclei (AGN) and Gamma-ray bursts (GRBs). The possibility of generating astrophysically relevant pair plasmas in laboratories by intense lasers has been demonstrated by extensive theoretical and experimental studies, opening the way to mimic some of the physics relevant to astrophysics in the laboratory. There are several fundamental quantum electrodynamic (QED) processes for pair production under strong fields, i.e., multiphoton Breit-Wheeler (mBW) \cite{Nikishov1967}, Bethe-Heitler (BH) and the trident process \cite{Heitler1954,Chen2009,Martinez2019}.

The BH process dominates the pair production in high-Z thick (width $> \mathrm{mm}$) solid targets. This process occurs when energetic electrons move across the strong field of the heavy nucleus. The source of energetic electrons can be the hot electrons from the target irradiated by an intense laser \cite{Chen2009,Chen2015,Liang1998,Liang2015}. They can also originate from an external relativistic electron beam sent to the target \cite{Sarri2015}. In the scheme with a laser, a high-energy ($>100\ \mathrm{J}$), relativistic ($a_0>1$, where $a_0$ is the normalized field amplitude) laser with a duration $\gtrsim {\mathrm{ps}}$ is generally considered to generate abundant hot electrons. The BH-type pair production has been demonstrated by experiments, but with a limited positron yield of $< 10^{12}$ \cite{Chen2015}. By using high-energy lasers ($\sim 10\ \mathrm{kJ}$), the created pair density is anticipated to reach $\sim 10^{16}\ \mathrm{cm}^{-3}$ according to a scaling study \cite{Chen2015}. This pair yield is still low compared with the solid target (pair density fraction $\sim 10^{-7}$ of the target density), and the pair population is believed to have negligible impacts on the laser coupling with the target.
 
 The mBW process dominates the pair production from high-energy photons which propagate in ultraintense electromagnetic fields. An efficient mBW scheme is to initiate the QED cascade \cite{Bell2008, Fedotov2010, Bulanov2010, Nerush2011, Elkina2011, Gonoskov2013, Bashmakov2014, Gonoskov2015, Chang2015, Cascade-GoLP, Tamburini2017,Jirka2017,Luo2018} with a plasma seed to interact with two counterpropagating ultraintense lasers or multiple colliding lasers. Other schemes include the collisions between a laser and a leptonic beam \cite{Burke1997} or between two leptonic beams \cite{Chen1989,Fabrizio2019}. One practical setup for copious pair production in a laser facility is to use a single ultraintense laser to strike a solid target \cite{Ridgers2012}. This scheme is potentially feasible with the next-generation high-power laser facilities, e.g., ELI \cite{ELI} and EPAC \cite{EPAC}. Previous studies show this scheme is capable of producing a dense ($\gtrsim $ the relativistically-corrected critical density) pair plasma cushion in front of the target \cite{Kirk2013,Kostyukov2016,Yuan2018,Sorbo2018,Samsonov2019,Samsonov2020}. This dense pair production will change the local plasma profile and also the plasma response to the laser coupling. Therefore, our study on the self-consistent scenario of pair production [Sec. \ref{Sec: realistic_target}] is focused on this efficient scheme.
 
In the mBW scenario of a single laser striking a solid target, the formation of the pair plasma makes the laser directly interact with an electron-positron-ion mixed plasma with significant pair fraction. This kind of laser interaction with the mixed plasma has not yet been systematically explored, but it is critical in measuring, improving and tailoring the pair plasma generation. In the experiments which have been conducted or proposed for the pair production mentioned above, the currently available diagnostics for pair plasmas are remote detectors, such as magnetic spectrometers \cite{Chen2009,Sarri2015,Chen2015,Warwick2017,Liang2015}. There is difficulty in studying the local laser coupling into the mixed plasma via those detectors which mainly deliver the global information of the generated pair plasma, e.g., its energy spectra and inferred particle number. The well-established diagnostic using Thomson scattering \cite{Froula2006,Glenzer2009}, however, faces challenges when applied to the highly nonlinear interaction between an intense laser and a solid target.

In this paper, the laser interaction with an electron-positron-ion mixed plasma is studied, from the perspective of the associated high-order harmonic generation (HHG) \cite{Teubner2009,Thaury2010}. HHG is a frequency up-conversion process, and its spectrum reveals the local laser-driven plasma dynamics at the surface of the target or collective plasma motions inside the target. To our knowledge, HHG in the context of pair plasma generation has not been studied in detail. Here, we first focus on the laser interaction with an idealized mixed target, in order to obtain well-defined signals in the spectrum which enable concise and reliable analysis for the plasma dynamics based on the spectrum. This analysis gives insight into the distinct role of the pair population in the mixed plasma when coupled with the laser. The idealized target is assumed to have a sharp plasma-vacuum interface and uniform distributions of all the constituents, including the pre-formed pair plasma. The HHG spectrum is obtained by irradiating the target with a weakly relativistic laser. In addition to the idealized mixed target, we expand the study to the self-consistent scenario where the pair plasma is generated via the QED effects. The self-consistent scenario is focused on the scheme which uses a single laser to strike a solid target \cite{Ridgers2012}, because dense pair plasmas can be created, altering the plasma response to the laser coupling and the resultant HHG spectrum.
		
The spectra from the idealized target in extreme conditions (with extreme parameters, e.g., very high pair fraction or hot pair temperature), are examined, which help us to understand the simulations of self-consistent pair production. They show high-frequency modulations are present on the reflected laser field. These modulations enhance the high frequency components of the spectrum. Our study shows these modulations are produced by the direct laser coupling with the expanded pair plasma in front of the target. The features of the high-frequency modulations and enhanced high frequency radiation indicate a transition to the spectrum in the self-consistent scenario which also manifests enhanced high frequency radiation due to the laser interaction with the created highly dense and hot pair plasmas.

This paper is organized as follows. In Sec. \ref{Sec: idealized_target}, the HHG spectrum from a laser interaction with an idealized electron-positron-ion target is studied. Both theoretical analysis [Sec. \ref{Subsec: theory}] and particle-in-cell (PIC) simulations [Sec. \ref{Subsec: 1D_PIC_simulations}] are performed to study the plasma dynamics and the associated harmonic spectrum. Furthermore, the finite temperature of the pair plasma, and other related parameters are examined in Sec. \ref{Subsec: additional_effects}. In Sec. \ref{Sec: realistic_target}, the HHG study is expanded to the self-consistent scenario by performing QED-PIC simulations where the pair plasma is produced by an ultraintense laser striking a solid target. In Sec. \ref{Sec: discussion_conclusion}, the conclusions of the study in this paper are stated.

\section{HHG from an idealized electron-positron-ion mixed plasma}
\label{Sec: idealized_target}

\subsection{Theoretical analysis}
\label{Subsec: theory}

In our study, the idealized electron-positron-ion mixed target is assumed to be an overdense plasma with a pre-formed pair plasma. The charge neutrality is satisfied, i.e., $n_e=n_p+Z_in_i$, where $n_e$, $n_p$ and $n_i$ are the electron, positron and ion densities, respectively, and $Z_i$ is the ion charge state. For simplicity, this target is assumed to have immobile ions, a sharp interface ($L_n\ll \lambda _0$, where $L_n=n/\nabla n$ is the gradient length of plasma density and $\lambda_0$ is the laser wavelength) and uniform density distribution. The positron fraction is defined as $\alpha=n_p/n_e$. A $p$-polarized, weakly relativistic ($a_0\lesssim 1$) laser is incident with angle $\theta$, where $a_0$ is the normalized field amplitude. The boost technique \cite{Bourdier1983,Bulanov1994,Lichters1996} is employed here, enabling 1-dimensional (1D) studies on oblique laser incidence. Appendix \ref{Appendix: boost_technique} gives related notations and formalisms of this technique.
 
We use the well-established oscillating mirror model (OMM) \cite{Lichters1996,Teubner2003,Baeva2006,Teubner2009,Thaury2010,Rodel2012,Debayle2015} to analyze the plasma dynamics around the surface, and we neglect other mechanisms \cite{May2011} for the particle acceleration. Because the pair plasma yield in experiments is so far small ($\alpha \ll 1$) \cite{Sarri2015,Chen2015,Liang2015}, the electron population governs the mirror oscillation, and the positrons can be treated as test particles. For convenience, the theoretical analysis and the simulations are presented in the moving frame. In the following equations, the time, space and electric field are normalized by $\omega_{0}^{-1}$, $c/\omega_{0}$ and $m_e\omega_{0}c/e$, respectively. Here, $c$ is the speed of light in vacuum; $e$ and $m_e$ are the charge and mass of an electron; $\omega_{0}$ is the laser frequency. The governing equation for electron motion is given by \cite{Lichters1996}:
\begin{equation}
\label{Eq: electron_motion}
\frac{d^2X_e}{dt^2}=-\frac{\Gamma_d^3\omega_{pe}^2}{\gamma_{e}}X_e+\frac{a^2-a\tan \theta}{\gamma_{e}^2d_s}
\end{equation}
Here, $X_e$ is the electron displacement with respect to the initial surface (at $x=0$); $a$ is the driving field for plasma motion; $\gamma_{e}=\sqrt{[1+(a-\tan \theta)^2]/(1-\beta _{xe}^2)}$ is the Lorentz factor of electron motion and $\beta_{xe}=dX_{e}/dt$. In the non-relativistic limit ($a_0\ll 1$), $\beta_{xe}\ll 1$, $\gamma_{e}\simeq \Gamma_d$ where $\Gamma_d=\cos^{-1} \theta$ (see Appendix \ref{Appendix: boost_technique}). The leading $\omega_{0}$-frequency component of the laser field is used to drive the plasma motion \cite{Lichters1996}. Inside the target the driving field is formulated as $a\simeq a_s\exp (-x/d_s)\sin t$ for $x\geqslant 0$, where $d_s=\omega_{pe}^{-1}(\Gamma_d^2-{\omega_{pe}^{-2}})^{-1/2}$ is the inertial length, and $a_s=2a_0\omega_{pe}^{-1}[(1-\Gamma_d^{-2}{\omega_{pe}^{-2}})/(\cos ^2{\theta} -\omega_{pe}^{-2}\cos 2\theta)]^{1/2}$ is the amplitude at surface. Here, $\omega_{pe}=\Gamma_d^{-3/2}(n_{e}e^2/m_e\varepsilon_0)^{1/2}$ is the plasma frequency in the moving frame (see Appendix \ref{Appendix: boost_technique}) and $\varepsilon_0$ is the permittivity of free space. For overdense plasmas where $\omega_{pe} \gg \omega_{0}$, $d_s\simeq \Gamma_d^{-1}\omega_{pe}^{-1}$ and $a_s\simeq 2a_0{\Gamma_d}\omega_{pe}^{-1}$. 
 \begin{figure*}
	\centering
	\includegraphics[width=15.24cm, height=4.8cm]{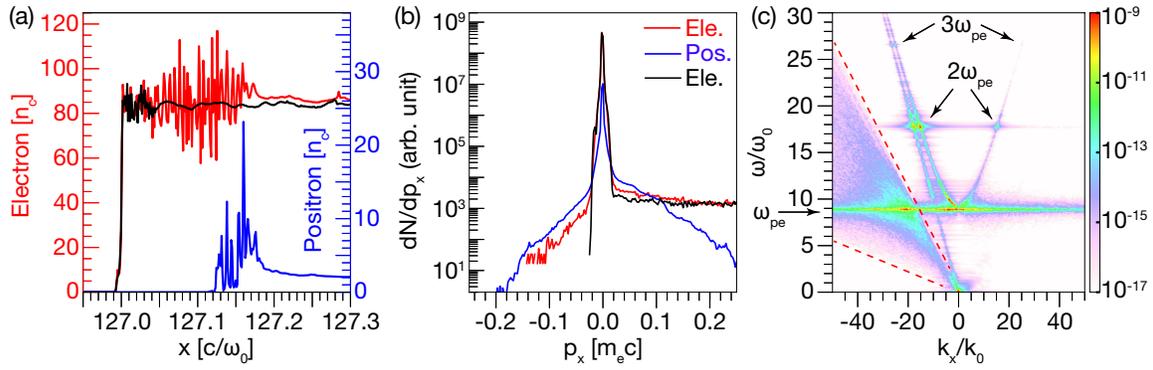}	
	\caption{(Color online). Results from a 1D PIC simulation with $n_i=84n_c$, $n_p=2n_c$ and $\theta=\pi/12$. (a) Plasma densities at $t=27.6\omega_0^{-1}$. The red and blue lines depict the electron and positron in the mixed target. The black line depicts the electron in the comparison electron-ion target. (b) Distributions of the longitudinal momentum $p_x$ of the particles inside the target after the laser-plasma interaction (at $t=129\omega_0^{-1}$), including the electrons (red) and positrons (blue) from the mixed target, and the electrons (black) from the electron-ion target. (c) $(\omega,k_x)$ power spectrum for the field $E_y$ from within the mixed target and from the right vacuum. Here, $k_0=\omega_{0}/c$. The dashed red lines indicate the region of plasma modes excited by hot electrons.}
	\label{fig: fig1}
\end{figure*}

When the laser field displaces the electrons, an electrostatic field is established as a restoring force dragging the electrons back. When the laser is obliquely incident, ``vacuum heating" \cite{Brunel1987,Gibbon1992} occurs and dominates over the ponderomotive force in the non-relativistic limit. This allows us to estimate the electrostatic field by balancing the Lorentz force and the restoring force, i.e., $E_{x0}\simeq -2a_0(\sin \theta/\cos^2\theta) \sin t$. Because the laser field exponentially decays at the surface, the generalized form for the space-dependent electric field is $E_x(x,t)\simeq E_{x0}\exp (-x/d_s)$. The equation of motion for the test positron is then given by:
\begin{equation}
\label{Eq: positron_motion}
\frac{d^2X_p}{dt^2}=\frac{E_x(X_p,t)}{\gamma_{p}}+\frac{a^2(X_p,t)+a(X_p,t)\tan \theta}{\gamma_{p}^2d_s}
\end{equation}
Similar to Eq. \eqref{Eq: electron_motion}, $X_p$ here is the positron displacement, $\gamma_{p}=\sqrt{[1+(-a-\tan \theta)^2]/(1-\beta _{xp}^2)}\simeq \Gamma_d$, where $\beta_{xp}=dX_{p}/dt \ll 1$. This equation is nonlinear, but its analytical solution can be obtained with few assumptions. Because the electrostatic force and the Lorentz force acting on the positron largely cancel each other, Eq. \eqref{Eq: positron_motion} can be approximated as $d^2X_p/dt^2\simeq a^2(X_p,t)/\gamma_{p}^2d_s$, indicating that the positrons are continuously pushed inward by the ponderomotive force of the laser. Furthermore, the field inside the target is assumed to be $a(x,t)\approx a_s\sin^2(\pi t/\tau _0)\sin t$, where the field's spatial dependence is neglected, because the equation of motion is solved within the inertial length region, i.e., $X_p<d_s$. Finally, $sin^2$ is chosen as the laser's temporal profile ($\tau_0$ is the pulse duration). The analytical solution to the approximated positron equation, in the small $t$ regime ($t/\tau_0\ll 1$), is given by $X_p=\kappa\tau_0^{-4}t^6$, where $\kappa=\pi ^4\Gamma_da_0^2\omega_{pe}^{-1}/15$. This solution should be truncated at $X_p\sim d_s$. Therefore, the maximum bulk velocity of the positrons is given by:
\begin{equation}
\label{Eq: positron_velocity}
v_p/c\simeq 2.41\Gamma_d^{5/6}a_0^{1/3}(n_e/n_c)^{-1/2}(\tau_0/2\pi )^{-2/3}
\end{equation}
where $n_c=m_e\varepsilon _0 \omega_0^2/e^2$ is the critical density. We have confirmed that Eq. \eqref{Eq: positron_velocity} provides a good estimation in the small $\alpha $ regime, by both numerically solving Eq. \eqref{Eq: positron_motion} and via PIC simulations with different $n_e$ and $\theta$. The more complex plasma dynamics with highly kinetic effects occurring at larger $\alpha$ will be explored elsewhere.

Plasma radiation at $\omega_{pe}$ has been previously identified from the plasma waves solely and impulsively driven by attosecond hot electron bunches produced in periodic vacuum heating \cite{Sheng2005,Quere2006,Borot2012,Kahaly2013,Yeung2013,Thaury2010,Boyd2000,Rovira2012}. In the mixed target, however, the positron plays a significant role and drastically changes the plasma wave excitation and the radiation features. The continuous positron acceleration [Eqs. \eqref{Eq: positron_motion} and \eqref{Eq: positron_velocity}] forms a dense positron beam which streams into the plasma. This beam drives a perturbation in the background electrons, where the electrons are attracted by the positive driver; large-amplitude plasma oscillations are therefore generated (shown later). Inside the target, where the laser field vanishes, the positron beam starts to stretch under the electrostatic field of the background plasma wave. A significant number of positrons (about one half), are pulled back by this field, and thus carry negative momenta. This reflux of positrons induces strong plasma waves moving backward to the front side of the target. Counterpropagating plasma waves are therefore created producing $2\omega_{pe}$ (and even $3\omega_{pe}$)-radiation via inverse two-plasmon decay \cite{Smith1971,Vasquez2002,Lichters1997,Kunzl2003}. This is also qualitatively different from the conventional scenario \cite{Lichters1997,Gibbon1997,Teubner1997,Teubner1999,Boyd2000,Kunzl2003,Rovira2012} of $2\omega_{pe}$-radiation where the backward wave is driven by the hot electrons reflected from the rear surface of the target.

\subsection{Particle-in-cell (PIC) simulations}
\label{Subsec: 1D_PIC_simulations}

We have verified the above physical picture by performing 1D/2D PIC simulations with {\small{OSIRIS}} \cite{Fonseca2002}. A wide range of parameters are examined, including the dependences of $\alpha$, $n_i$ ($Z_i$ is set to be $1$ for simplicity), angle $\theta$ and pair temperature $T_p$. In the 1D simulations, the spatial and temporal resolutions are $\Delta x=1.8\times 10^{-4}\lambda_0$ and $\Delta t=1.6\times10^{-4}T_0$, where $T_0=2\pi \omega_{0}^{-1}$ is the laser period. The mixed target is placed at $127c/\omega_0 \leqslant  x\leqslant 190c/\omega_0$. A short-pulse laser is initialized in the box with a profile of $a_0\sin^2(\pi t/\tau _0)$ where $a_0=0.5$ and $\tau_0=20T_0$. Meanwhile, simulations with pure electron-ion plasmas are performed for comparison, to verify that the observed physics is due to the positron population. Convergence tests with higher resolutions have been performed showing the same results as presented here. 

Figure \ref{fig: fig1} shows results from a simulation with $n_i=84n_c$ and $\alpha =2.3\%$. Large-amplitude short-wavelength plasma oscillations are excited by the positron beam [Fig. \ref{fig: fig1}(a)]. The positrons pile up while being pushed inwards by the laser, explaining the appearance of high positron density ($n_p\sim 23n_c$). The density of the trailing electrons is $n_e\sim 110n_c$ $(\simeq n_i+n_p)$, indicating the perturbation of electron density is roughly the same as the positron driver. In contrast, the maximum density perturbation in the pure electron-ion target is only $\sim 7n_c$ close to the surface. The electrons undergo a density modulation with a wavelength of $\sim 0.007c/\omega_0$ which corresponds to the plasma wavelength, i.e., $\lambda _p\simeq 2\pi v_p/\omega_{pe}$, where $\omega_{pe}=8.8\omega_{0}$ and the bulk positron velocity at this moment is $\sim 0.01c$, indicating the onset of beam-plasma instability \cite{Boyd2003}. The positrons are accelerated to a peak bulk velocity of $0.03c$ by $t=50\omega_{0}^{-1}$, in agreement with Eq. \eqref{Eq: positron_velocity}. The positron beam starts to stretch in velocity space when travelling across the excited plasma waves. Eventually, some positrons are further accelerated with velocities reaching $0.3c$, while $\sim 50\%$ of the positrons carry negative momenta forming a positron reflux [Fig. \ref{fig: fig1}(b)]. The positron reflux shows a high velocity tail reaching $ \sim -0.2c$, and a significant number of electrons also carry negative momenta with a tail velocity $\sim -0.14c$, resembling the positron driver. In the pure electron-ion target, however, the reflux of electrons is negligible. 

The $E_y$ power spectrum in the reciprocal space $(\omega,k_x)$ is shown in Fig. \ref{fig: fig1}(c). The spectrum with $k_x<0$ (or $k_x>0$) represents the radiation/plasma modes in transmission (or reflection). The transmitted radiation in the right-hand side vacuum is captured by the inclined line at $\omega=-ck_x$. The radiation propagating inside the drifting plasma is indicated by the bright parabolic dispersion contour. The excited forward and backward plasma modes are clearly shown by the line at $\omega \simeq  \omega_{pe}$. The total power of the backward plasma waves in the mixed target, calculated by summing all the modes with $k_x>0$ and $\omega \simeq\omega_{pe}$ in Fig. \ref{fig: fig1}(c), is $\sim 40$ times higher than the counterpart power from the electron-ion target. The coupling between these counterpropagating plasma waves produces strong $2\omega_{pe}$ ($3\omega_{pe}$)-radiation present in both transmission and reflection. The dashed red lines indicate the plasma modes driven by the hot electron bunches whose velocities range within $0.1c\sim 0.6c$, consistent with the slopes of dashed lines. 

\begin{figure}[t]
	\includegraphics[width=8.5cm, height=9.11cm]{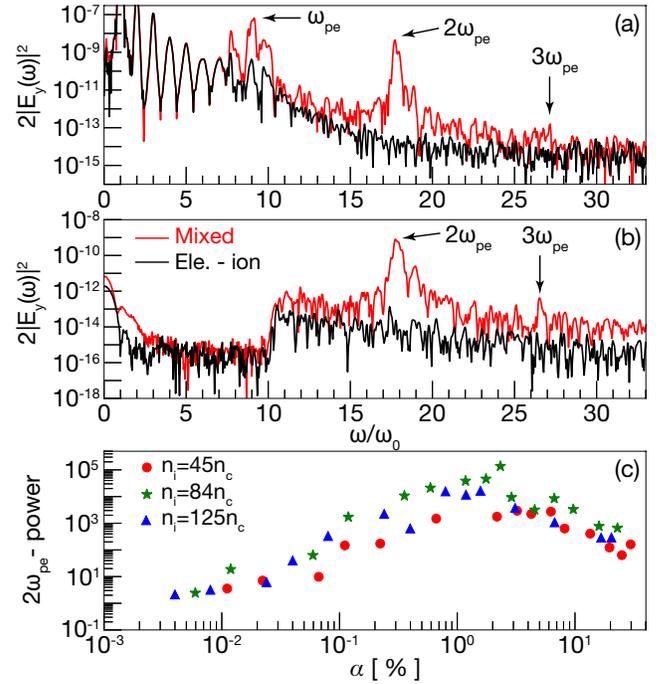}
	\caption{(Color online). Time-integrated power spectrum of $E_y$ in reflection (a) and in transmission (b) from the simulation in Fig. \ref{fig: fig1}. The red and black lines are for the mixed and electron-ion targets, respectively. (c) The relative power of $2\omega_{pe}$-radiation (normalized by the power at $2\omega_{pe}$ in the electron-ion target) in reflection as a function of positron fraction $\alpha$. The red, green and blue symbols are for $n_i=45n_c$, $84n_c$ and $125n_c$, respectively.}
	\label{fig: fig2}
\end{figure}

The time-integrated spectra via fast Fourier transform (FFT) in reflection/transmission, i.e., $E_y(\omega)=\mathrm{FFT}[E_y(x_0,t)]$, where $x_0$ is the fixed position of the observer placed in the left-hand/right-hand vacuum, are shown in Figs. \ref{fig: fig2}(a) and \ref{fig: fig2}(b). Prominent $n\omega_{pe}$-signals are observed in the reflection spectrum for the mixed target, in contrast to the electron-ion target. The $\omega_{pe}$- and $2\omega_{pe}$-radiation amplitudes are $155$ and $1.4\times 10^5$ times higher than in the electron-ion target. For the transmission spectrum, well-defined high order plasma radiation ($2\omega_{pe}$ and $3\omega_{pe}$) are also observed, with amplitudes $6000$ and $200$ times higher than in the electron-ion target.

The scaling study shows that the $2\omega_{pe}$-radiation peak is pervasive over broad ranges of $\alpha$ and $n_i$ [see Fig. \ref{fig: fig2}(c)]. The optimal positron fraction is $\alpha \sim 2\%$ at which the $2\omega_{pe}$-radiation power is maximized. The maximum $2\omega_{pe}$-radiation enhancement exceeds $5$ orders of magnitude as shown in Fig. \ref{fig: fig2}(a). The beam-plasma instability generally depends on the beam density $n_p$ and its relative velocity spread $\Delta v_p/v_p$ \cite{Rovira2012,Boyd2003}, because the linear growth time for the wave excitation is $t_g\propto n_p^{-1}(\Delta v_p/v_p)^2n_e^{1/2}$ and the bandwidth of the excited wave scales as $\Delta \omega/\omega_{0}\sim (\Delta v_p/v_p)(n_e/n_c)^{1/2}$. In the low $\alpha$ regime, the positron beam is formed with a low temperature/emittance, i.e., small dispersion in both coordinate and velocity spaces, producing narrow $n\omega_{pe}$ spectral peaks. The $2\omega_{pe}$-radiation power in this regime increases with an approximate power-law dependence on $\alpha$ [see Fig. \ref{fig: fig2}(c) for $\alpha \lesssim 2\%$]. The lowest positron density, which can result in a notable $2\omega_{pe}$-peak, is $n_p\sim 0.005n_c$. In the high $\alpha$ regime, the bulk positron velocity decreases with $\alpha$. The pile-up of the laser-pushed positrons will lead to the slowdown or even breakdown of the positron beam, significantly increasing its temperature/emittance. The bandwidth of the excited plasma waves is broadened, resulting in the drop and eventual saturation of the $2\omega_{pe}$-radiation power.

We highlight the spectrum in the very high $\alpha$ regime ($\alpha > 0.1$) in Fig. \ref{fig: fig3}. The $2\omega_{pe}$-radiation is still identified at $\sim 19\omega_{0}$, but broadband radiation, especially between $\omega_{pe}$ and $2\omega_{pe}$, is observed, flattening the $n\omega_{pe}$-radiation peaks. This reveals broadband plasma waves are excited by the hot electrons, and the dense positron beam which has large coordinate/velocity spreads. It is worthy noting that the plasma (positrons and electrons) expansion into the vacuum becomes significant when $\alpha$ is very high. The laser directly interacts with the expanded pair plasma, imposing high-frequency modulations on the reflected laser field. These modulations give rise to the enhanced high frequency components ($\gtrsim 3\omega_{pe}$) in the spectrum. We note the spectral feature in the very high $\alpha$ regime, i.e., broadband plasma radiation and enhanced high frequency harmonic components, are consistent with the self-consistent scenario [Sec. \ref{Sec: realistic_target}] where denser and hot pairs are produced in the interaction region.

\subsection{Additional effects impacting the HHG spectrum}
\label{Subsec: additional_effects}
The plasma radiation identified in the idealized mixed plasma has been found to be robust, even when including additional effects. First, significant $2\omega_{pe}$-radiation is generally observed for oblique incidence with $\theta <60^{\circ}$ as predicted in Ref. \cite{Kunzl2003}. Second, with $s$-polarized lasers, the $2\omega_{pe}$-radiation is missing, because the energetic hot electron bunches are absent and thus no forward plasma waves are excited. Without the strong field of the background plasma wave, the spread of the positron beam is weak and the positron reflux does not occur. Third, we have examined the effect of finite pair temperature $T_p$ [Fig. \ref{fig: fig4}(a)]. This enables a HHG study which accounts for the thermal effects and pair plasma expansion, creating a situation closer to the self-consistent scenario where the pair is produced by the ultraintense laser plasma interaction. In the low $T_p$ regime, where the thermal velocity ($v_{th}=(T_p/m_ec^2)^{1/2}$) is smaller than the positron velocity given by Eq. \eqref{Eq: positron_velocity}, the well-defined positron beam is formed as before, but the power of $2\omega_{pe}$-radiation drops with $T_p$ due to the thermal spread of the beam. For the parameters used in Fig. \ref{fig: fig4}(a), this low $T_p$ regime corresponds to $T_p\lesssim 0.5 \ \mathrm{keV}$, consistent with the scaling study in Fig. \ref{fig: fig4}(a). In the high $T_p$ regime, i.e., $T_p \gtrsim \textrm{keV}$, multiple positron bunches with high velocities are observed, in contrast to the single positron beam in the low $T_p$ regime. These positron bunches also produce $2\omega_{pe}$-radiation. Our simulations show the well-defined $2\omega_{pe}$-radiation peak survives for $T_p\lesssim 30\ \mathrm{keV}$.

  \begin{figure}
	\includegraphics[width=8.5cm, height=4.25cm]{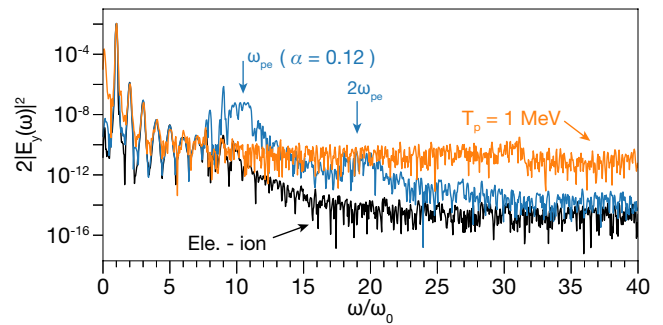}
	\caption{(Color online). HHG spectrum in reflection with extreme parameters. The blue line is with a very high positron fraction of $\alpha = 0.12$. Other parameters are the same as in Fig. \ref{fig: fig1}. The orange line is with a hot pair temperature of $T_p = 1 \ \mathrm{MeV}$. Other parameters are the same as in Fig. \ref{fig: fig4}(a). The black line represents the electron-ion target as a comparison.}
	\label{fig: fig3}
\end{figure}

\begin{figure}
	\includegraphics[width=8.5cm, height=5.83cm]{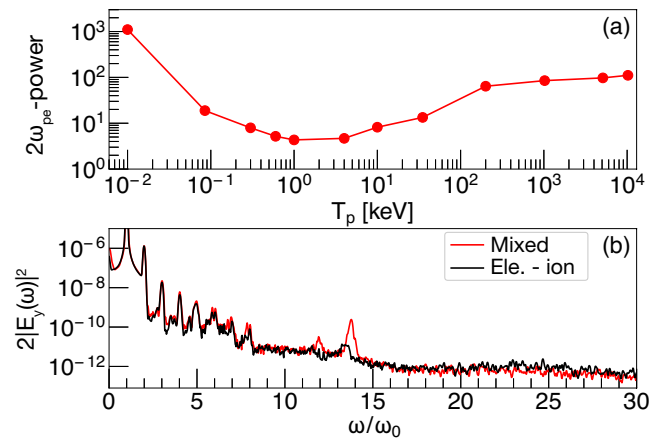}
	\caption{(Color online). (a) The $2\omega_{pe}$-radiation (normalized in the same way as Fig. \ref{fig: fig2}(c)) in reflection as a function of $T_p$. The initial positron density is $n_{p}=0.5n_c$ and other parameters are the same as Fig. \ref{fig: fig1}. (b) Spectrum of the spatial FFT at the end of the interaction, for the reflected field in a 2D simulation with mobile ions, $n_i=45n_c$, $n_p=2n_c$, $T_p=0$ and $\theta =\pi /4$. The laser is the same with Fig. \ref{fig: fig1}, but with a Gaussian-type finite spot size.}
	\label{fig: fig4}
\end{figure}

 \begin{figure*}
	\includegraphics[width=18.5cm, height=6.29cm]{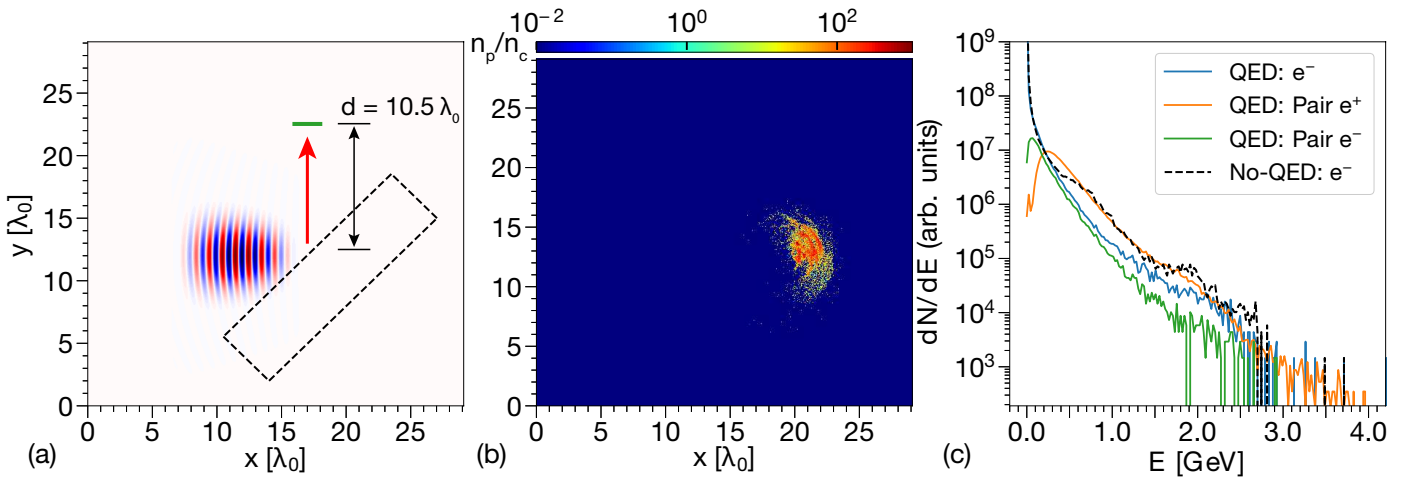}
	\caption{(Color online). Results from a 2D QED-PIC simulation with HHG diagnostics. (a) Schematic plot showing the simulation configuration (see the text for details), including the drive laser (red/blue colors) and target (dashed rectangle). The red arrow shows the ideally specular light reflection, given the target acts as an unperturbed mirror. An observer (horizontal green segment) is deployed to record fields for spectrum analysis. The vertical distance of the observer from the target is $d=10.5\lambda_0$. (b) Normalized density profile of the created positrons ($n_p/n_c$) at $t=11.4T_0$. The maximum positron density is $>2000n_c$. (c) Energy spectra at $t=13.4T_0$ from the QED simulation for the target electron (blue), created pair positron (orange) and electron (green), and from the non-QED simulation for target electron (dashed, black).}
	\label{fig: fig5}
\end{figure*}

When the pair temperature is even higher, i.e., $T_p\sim \textrm{MeV}$, the spectrum shows new features, as shown in Fig. \ref{fig: fig3} for one example with $T_p=1 \ \mathrm{MeV}$. Broadband plasma waves are excited due to the relativistic thermal spread of the particle beams. The $n\omega_{pe}$-radiation peaks are completely smeared out. In addition, the simulation shows the thermal expansion dominates the interaction, leading to a hot pair-dominated plasma in front of the target, consistent with previous studies \cite{Liang1998,Chen2015}. The laser coupling with the pair plasmas induces high-frequency modulations on the reflected laser field. These modulations enhance the power of high frequency harmonics ($\gtrsim 3\omega_{pe}$) in the spectrum. This results in a broadband plateau in the spectrum. The height of the plateau increases with $T_p$, explaining the rise of radiation power at $2\omega_{pe}$ in the $\sim \textrm{MeV}$ regime in Fig. \ref{fig: fig4}(a). We note that the spectral features here with $\sim \ \mathrm{MeV}$ pair plasma are also consistent with the self-consistent scenario studied in Sec. \ref{Sec: realistic_target}, where denser and hotter pair plasmas are created in the interaction region.

We have so far assumed the ions form a fixed background, because their motion is generally negligible in the femtosecond and moderately relativistic laser-plasma interaction \cite{Rovira2012}. After performing simulations with mobile ions, we find that in the pure electron-ion target, the ion motion can play a role in the $2\omega_{pe}$-radiation; however, in the mixed target, the positrons dominate the positive charge dynamics and excitations of the counterpropagating plasma waves, and no evidence of the ion motion impact has been identified. This is also confirmed in multi-dimensional simulations with mobile ions [Fig. \ref{fig: fig4}(b)], where the power of the $2\omega_{pe}$-radiation is $>20$ times higher than in the electron-ion target. 

\section{HHG in self-consistent ultraintense laser scenario}
\label{Sec: realistic_target}
In this section, we expand the HHG study to a self-consistent scenario of pair production. The study is focused on using an ultraintense laser to strike a solid target \cite{Ridgers2012}, because this scheme is shown to be capable of generating abundant pair plasmas resulting in a mixed target with significant pair fraction ($\alpha \lesssim 1$). We anticipate the corresponding HHG spectrum to show the distinct plasma response to the laser coupling when dense pair plasmas are present. Various effects will occur in this scenario, e.g., relativistic particle acceleration and target expansion/deformation, and these effects have significant impacts on the HHG spectrum \cite{Chopineau2019}. To have a self-consistent study on this scenario, we resort to the QED-PIC simulation with {\small{OSIRIS}} \cite{Fonseca2002} which accounts for the photon emission and pair production via the mBW process.

The simulation set up is shown in Fig. \ref{fig: fig5}(a). The drive laser, with $\lambda_0 =1\ \mu m$, is initialized inside the simulation box along $x$ direction. A fully ionized aluminum target, with realistic density $n_e=700n_c$, is deployed ensuring an oblique incidence with $\theta = 45^\circ$. The simulation box is $29\lambda_0\times 31\lambda_0$ with numerical resolutions of $\Delta x=\Delta y=0.0024\lambda_0$ and $\Delta t=0.0013T_0$. The drive laser is circularly polarized, with $I_0= 5\times 10^{24}\ \mathrm{Wcm^{-2}}$, $sin^2$-temporal profile, $\tau_0=10T_0$ and transverse focal spot (FWHM in field) of $3\lambda_0$. Similarly, we also perform the non-QED simulations, with the same parameters but without the QED effects, for comparison.

\begin{figure}
	\includegraphics[width=8.5cm, height=4.25cm]{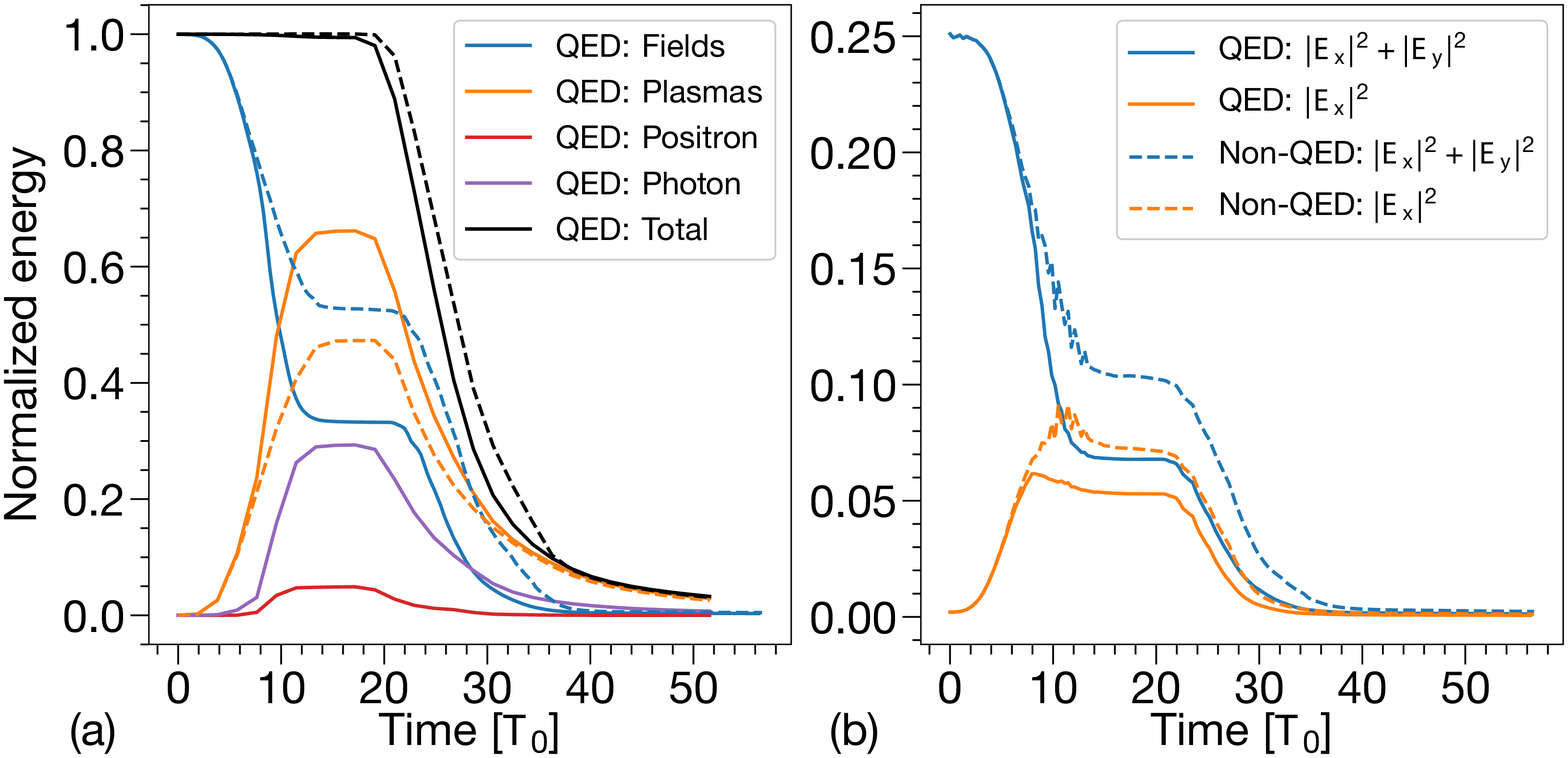}
	\caption{(Color online). System energy evolution, normalized by the initial total field energy, from the QED (solid) and non-QED (dashed) simulations [Fig. \ref{fig: fig5}]. (a) The blue lines show the field energy (electric $+$ magnetic), and orange lines show the plasma energy (background plasma $+$ pairs $+$ photons). The black lines show the system energy (field energy $+$ plasma energy). The purple/red lines specifically show the photon/positron energies, respectively. (b) Energy of the electric field in ($x,\ y$) plane: blue lines show $|E_x|^2+|E_y|^2$, and orange lines specifically show $|E_x|^2$.}
	\label{fig: fig6}
\end{figure}

Significant pair production (pair density $\textgreater 100n_c$) begins at $t\simeq 7T_0$, and this timing is also verified by the positron energy evolution in Fig. \ref{fig: fig6}. The positron profile is shown in Fig. \ref{fig: fig5}(b) at $t=11.4T_0$. The maximum positron density exceeds $2000n_c\sim a_0n_c$, consistent with previous studies \cite{Kirk2013,Nerush2015,Kostyukov2016,Yuan2018,Sorbo2018,Samsonov2019}. The energy spectra of the plasmas are shown in Fig. \ref{fig: fig5}(c). Although the laser-plasma coupling is highly nonlinear, the spectra of pair species show thermal tails whose temperatures are $\sim 200 \ \mathrm{MeV}$ based on exponential fits. The system energy is well conserved [black lines in Fig. \ref{fig: fig6}(a)]. The system energy drops after $t\sim 20T_0$, because the laser and energetic particles start to escape the simulation box. The laser coupling into the plasma (including the emitted photons and pair particles) saturates since $t\sim 13T_0$, indicating a higher coupling efficiency of $66\%$ in the QED simulation than in the non-QED one (just $ 47\%$). The higher coupling efficiency is due to the QED effects, i.e., photon emission and pair production whose energies make up $\sim 44\%$ and $\sim 15\%$ of the plasma energy, respectively. We pay special attention to the electric fields polarized in the ($x,\ y$) plane [Fig. \ref{fig: fig6}(b)]. For the laser-plasma configuration here, the ideally specular reflection is along $+y$ direction shown by the red arrow in Fig. \ref{fig: fig5}(a). Therefore, $E_x$ is the main component of the reflected field, and its energy in the QED simulation is just $74\%$ of the non-QED simulation, also indicating the higher absorption (lower reflection) when the QED effects are enabled \cite{Kirk2013,Nerush2015,Sorbo2018}. Given that the laser is reflected along a wide range of directions due to the deformation of the target surface by laser's radiation pressure \cite{Wilks1992}, the total electric energy ($|E_x|^2+|E_y|^2$) is also examined. It shows that $73\%$ of the electric energy is absorbed by the plasma in the QED simulation, while only $58\%$ is absorbed in the non-QED simulation. The difference in laser absorption is further verified by the corresponding HHG spectrum shown below.

  \begin{figure}
	\includegraphics[width=8.5cm, height=7.56cm]{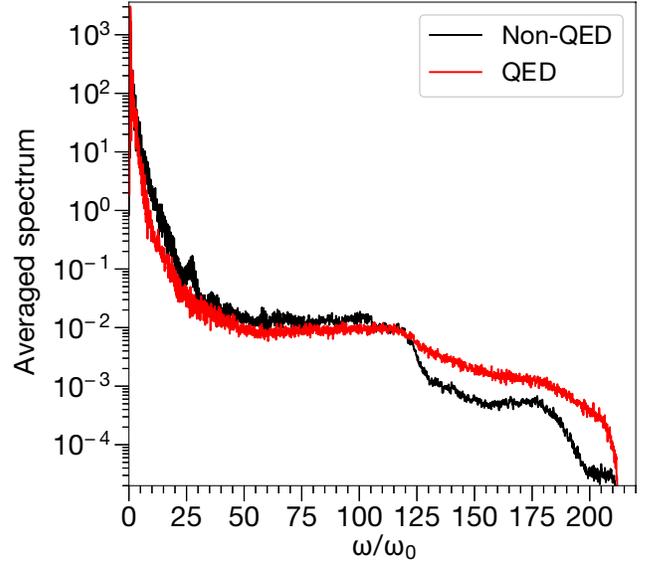}
	\caption{(Color online). Spatially averaged power spectrum of the reflected laser from the QED (red) and non-QED (black) simulations, respectively. The spectrum is obtained from the observer shown in Fig. \ref{fig: fig5}(a), and defined as $\overline{E_x(\omega)}=\sum \limits_{x_0=x_1}^{x_0=x_2}2|\mathrm{FFT}[E_x(x_0,t)]|^2$, where $(x_1,x_2)$ is the spanning region of the detector, with $x_2-x_1=2\lambda_0\sim $ laser spot size, over which the spectrum is averaged.}
	\label{fig: fig7}
\end{figure}

Figure \ref{fig: fig7} compares the spatially averaged, time-integrated spectrum from the observer depicted in Fig. \ref{fig: fig5}(a) of the QED and non-QED simulations. There are two features visible in the spectra. First, the low order harmonics of the non-QED simulation are of higher magnitudes than the QED counterpart. This indicates a higher laser reflection when the QED processes are disabled, consistent with the energy comparison in Fig. \ref{fig: fig6}. Second, broadband spectra of the high frequency harmonics are present in the QED simulation, in comparison to weaker counterparts in the non-QED spectrum. Corresponding to the first feature, the fundamental mode in the spectrum shows a stronger red-shift and higher power in the non-QED simulation than in the QED counterpart, because the hole-boring velocity is faster in the non-QED simulation due to its higher laser reflection.

The field ($E_x$) profiles (not shown here), recorded by the same observer, begin to differ between the QED and non-QED simulations at $t\simeq 17T_0$. Two differences are identified, i.e., the field amplitude is weaker in the QED simulation than in the non-QED simulation, and high-frequency modulations are present in the field from QED simulation but absent from the non-QED counterpart. The high-frequency modulations on the field are found to have frequencies $\gtrsim 110\omega_{0}$ and they therefore enhance the high frequency harmonics in the QED spectrum. These two differences respectively corroborate the two features summarized for the spectrum in Fig. \ref{fig: fig7}. When significant pair production is initiated (at $t\simeq 7T_0$), the laser field reflected by the target arrives at the observer at $t\sim 17T_0$ after propagating in vacuum for $d = 10.5\lambda_0$ [as shown Fig. \ref{fig: fig5}(a)]. The laser arrival coincides with the starting time (mentioned before) after which the recorded fields begin to differ. Therefore, the differences of recorded fields are attributed to the pair production, and this can be used to estimate the onset of pair production. 

On the one hand, the QED effects deplete the laser energy, resulting in weaker laser reflection and thus weaker low order harmonics in the spectrum. On the other hand, the laser interaction with the dense ($\gtrsim a_0n_c$) and hot ($\sim 200\  \textrm{MeV}$) pair plasma produces high-frequency modulations on the reflected laser. These modulations are recorded by the observer (mentioned above), and they enhance the high frequency components in the QED spectrum in Fig. \ref{fig: fig7}. The enhancement of the high frequency harmonics and formation of a high-frequency broadband plateau in the spectrum are analogous to the spectral features identified from an idealized mixed target in extreme conditions [see Fig. \ref{fig: fig3}] where the high frequency harmonics are also enhanced due to the direct laser interaction with the pair-dominated plasmas in front of the target.

The ultraintense laser-plasma interaction introduces high inhomogeneity (densities and momenta) in both space and time, to the mixed plasma. These inhomogeneities lead to the excitement of broadband plasma waves. The simulations suggest instabilities, analogous to the oblique filamentation instability \cite{Bret2005,Gremillet2007,Ghizzo2020a}, are observed. The HHG spectrum further depends on the solid angle where the observer is deployed, because the target surface is observed to be deformed in the simulations, by both the hole-boring process and the Rayleigh-Taylor like instability \cite{Pegoraro2007,Wu2014} at the surface. This will be explored in future publications.

\section{Conclusions}
\label{Sec: discussion_conclusion}
As introduced in Sec. \ref{Sec: introduction}, both BH and mBW scenarios have been conceived to produce pair plasma using intense lasers. The BH scenario has been demonstrated in experiments, but the pair yield is low ($\alpha<10^{-7}$) as analysed before. Therefore, the pair production in the BH scenario is unlikely to have significant impact on the HHG spectrum according to our scaling study in Fig. \ref{fig: fig2}(c). Moreover, the numerical study on the BH scenario should account for multiple processes \cite{Martinez2019}, e.g., the BH process, particle collisions and Bremsstrahlung.
 
For the mBW scenario using a laser to strike a solid target, our results in Sec. \ref{Sec: realistic_target} show abundant photon/pair production. The spectrum is modified due to the significant QED effects. Previous studies show that a plasma cushion, whose density far exceeds $a_0n_c$, can be formed in front of the target \cite{Kirk2013,Kostyukov2016,Yuan2018,Sorbo2018,Samsonov2019}. The collective motion of the pair plasma, e.g., the propagation of the cascade front \cite{Samsonov2019,Samsonov2020}, is also observed. This provides an exciting micro-laboratory where a dense pair plasma can be studied through the associated HHG spectrum. The spectral features in these situations need further systematic studies which are beyond the scope of this paper.

In conclusion, the HHG spectrum from the laser interaction with an electron-positron-ion mixed plasma is studied. In the idealized mixed plasma with a sharp initial plasma-vacuum interface and uniform density distribution, prominent $n\omega_{pe}$-radiation peaks are observed in the spectrum. These signals are produced by the strong counterpropagating monochromatic plasma waves via the inverse two-plasmon decay. The dense laser-accelerated positron beam acts as the driver exciting the counterpropagating waves. 1D PIC simulations show the $2\omega_{pe}$-radiation is discernible with a positron fraction $\alpha$ as low as $\sim 10^{-5}$, and its amplitude at the optimal $\alpha$ ($\sim 2\%$) exceeds $5$ orders of magnitude higher than in the electron-ion comparison target. However, if an idealized target is in extreme conditions, i.e., the pair plasma is very dense (high pair fraction $\alpha>0.1$) or hot (relativistic temperature $\sim \mathrm{MeV}$), the resultant spectrum shows different features. In both conditions, broadband plasma waves are excited. Therefore, broadband plasma radiation is generated and the characteristic $n\omega_{pe}$-peaks are flattened (even completely smeared out). Furthermore, a pair-dominated plasma is formed by the pair plasma expansion in front of the target. The direct laser coupling with this pair-dominated plasma gives rise to high-frequency modulations on the reflected laser field. These modulations enhance the high frequency components ($\gtrsim 3\omega_{pe}$) of the spectrum. The spectral features in these two conditions show a transition to the spectrum in the self-consistent scenario [Sec. \ref{Sec: realistic_target}] which also manifests enhanced high frequency radiation due to the laser interaction with the created highly dense and hot pair plasmas.

With the parameters chosen in Sec. \ref{Sec: realistic_target}, our results show dense ($\gtrsim a_0n_c$), hot ($\sim 200\ \mathrm{MeV}$) and relativistic ($\sim \mathrm{GeV}$) pair plasmas are created. Furthermore, the target becomes highly inhomogeneous in both density and momentum within the interaction region. When the dense pair production begins, the reflected laser field is then found to be significantly depleted and have high-frequency modulations. On the one hand, the field depletion indicates a high laser absorption due to the QED effects, leading to weak power of the low order harmonics in the QED spectrum. On the other hand, high-frequency modulations are induced by the laser interaction with the created dense pair plasmas. The modulations result in enhanced high order harmonics in the QED spectrum in contrast to the non-QED spectrum where these field modulations are absent. Our study shows the HHG analysis has the potential acting as a side window into understanding the physics for pair production, e.g., the onset of the pair production and the distribution of pair density.

\begin{acknowledgments}
W.L.Z. gratefully acknowledges useful discussions with Y. X. Zhang, T. Silva, M. Pardal and B. Martinez. The comments of the anonymous referees are appreciated. This work was supported by the European Research Council (ERC-2015-AdG grant No. 695088), and FCT (Portugal) Grants No. SFRH/IF/01780/2013. We acknowledge PRACE for awarding us access to MareNostrum at Barcelona Supercomputing Center (BSC). Simulations were performed at the IST cluster (Lisbon, Portugal) and at MareNostrum (Spain).
\end{acknowledgments}

{\section{\appendixname}}
\begin{appendices}
	\section{The boost technique for the study on the oblique laser incidence}
	\label{Appendix: boost_technique}
\end{appendices}
The boost technique \cite{Lichters1996} enables 1-dimensional (1D) studies on oblique laser incidence. In the laboratory frame, the laser is incident with a $\theta$-angle respect to the target normal. In the boosted frame where the plasma drifts in the transverse direction at a velocity $v_d=-c\sin \theta$, the laser incidence becomes normal. Here, $c$ is the speed of light in vacuum. One can readily convert between these two reference frames using Lorentz transformations, i.e.,
\begin{subequations}
	\label{Eq: Lorentz_LM}
	\begin{eqnarray}
	n_{eM}=&\Gamma_d n_{eL} \\
	\omega_{0M}=&\Gamma_d ^{-1}\omega_{0L} \\
	k_{xM}=&k_{xL} \\
	k_{yM}=&0
	\end{eqnarray}
\end{subequations}	
 where the subscripts `L' and `M' stand for the laboratory frame and moving frame, respectively; '$x$' denotes the longitudinal (target normal) component; $\Gamma_d=(1-v_d^2/c^2)^{-1/2}=\cos^{-1} \theta$; $k_{xL} =k_{0L}\cos\theta$, where $\omega_{0}$ and $k_0$ are the laser frequency and wave number, respectively. While the Langmuir frequency remains invariant, the plasma frequency $\omega_{pe}$ has a Doppler shift between the two frames \cite{Tajima1979,Bulanov1994,Kunzl2003}, i.e., 
 \begin{equation}
 \omega_{peM}=\Gamma_d^{-1}\omega_{peL}
 \end{equation}
 Here, $\omega_{peL}=(n_{eL}e^2/m_e\varepsilon_0)^{1/2}$ and $\omega_{peM}=\Gamma_d^{-3/2}(n_{eM}e^2/m_e\varepsilon_0)^{1/2}$, where $e$ and $m_e$ are the charge and mass of an electron, and $\varepsilon_0$ is the permittivity of free space.

\newpage

\end{document}